\documentclass[12pt]{article}

\usepackage{a4wide}

\usepackage[latin1]{inputenc}
\usepackage{color}
\usepackage[colorlinks,linkcolor=black,citecolor=black]{hyperref}

\usepackage{verbatim,version}
\includeversion{ebook}
\includeversion{paper}

\usepackage{graphicx}
\usepackage{moreverb}

\usepackage{latexsym}

\usepackage{float,endfloat}
\floatstyle{ruled}
\restylefloat{table}
\restylefloat{figure}

\providecommand{\hhref}[2][]{\hyperref[#1]{#2}}

\providecommand{\Caption}[3][box]
   {\caption{\hhref[ret:#2]{#3}\label{#1:#2}}}
   
\newfloat{Box}{tp}{code}

\usepackage{alltt}
\usepackage{theorem}

\renewcommand{\abstract}[1]{\noindent\hrulefill

#1

\noindent\hrulefill\\}

\newcommand{\In}{{\bf in}}

\newcommand{\Let}{{\bf let}}

\newcommand{\Seq}[3][,]{\ensuremath{#2_1#1#2_2#1\cdots#1#2_{#3}}}

\theoremheaderfont{\rm}
\theoremstyle{plain}
\newtheorem{prop}{PROPOSITION}

\newenvironment{proof}{\begin{trivlist}\item PROOF. }{QED\end{trivlist}}

\newcommand{\lam}{\ensuremath{\lambda\,}}

\newcommand{\presentation}[1]{}

\newcommand{\args}[2][]{\{#2\}\subox{#1}}
\newcommand{\sigs}[2][]{[#2]\subox{#1}}
\newcommand{\is}{=}
\newcommand{\scope}[2][E]{\emph{#1}\subox{#2}}
\newcommand{\sub}[1]{\ensuremath{_{#1}}}
\newcommand{\subox}[1]{\ensuremath{_{\mbox{\tt\scriptsize #1}}}}
\newcommand{\ttsize}{}
\newcommand{\ttbox}[1]{\ensuremath{\mbox{\ttsize\tt #1}}}
\newcommand{\itbox}[1]{\ensuremath{\mbox{\it #1\/}}}
\newcommand{\embox}[1]{\ensuremath{\mbox{\em #1\/}}}

\newcommand{\Rule}[3][]{
  \begin{array}[t]{@{}l@{}}
    \TE{#2} % {\scope[R]{#2}}++
    \is
    #3 % {\scope[R]{#2}}b
    #1
  \end{array}
}

\newcommand{\TE}[2][\theta]{
  \ensuremath{\exists\subox{\,#2}#1.}
}
\newcommand{\Subst}[2][\theta]
{\ensuremath{\left[\rule{0mm}{0.25cm}#1\triangleleft#2\right]}}
\newcommand{\presume}[2][\\]
{#1 \dashv
\left[\rule{0mm}{0.3cm}%
\begin{array}[@{}c@{}]{l}\\[-2ex]
      #2
\end{array}\right]}

\newcommand{\defmac}[2]{
\ttbox{#1\,\ensuremath{\stackrel{\mathrm{\scriptsize def}}{=}}\,\{#2\}}
}

\renewcommand{\hhref}[2][]{#2}

\providecommand{\Caption}[3][box]
   {\caption{{#3}\label{#1:#2}}}

\includeversion{paper}
\excludeversion{ebook}
\newcommand{\smallsf}{}

\begin{document}

\begin{center}
{\Large\bf \hhref[ref:contents]{Programs as Proofs}}\\[3ex]
\begin{paper}
{\em Jørgen Steensgaard-Madsen\\
(Retired)}\\[3ex]
\end{paper}
\begin{ebook}\vspace{4ex}\end{ebook}
\textbf{Abstract}\\[1ex]
\begin{minipage}{0.9\textwidth}\small
The Curry-Howard correspondence is about a relationship between types and programs on the one hand and propositions and proofs on the other.  The implications for programming language design and program verification is an active field of research.
\\[2ex]
Transformer-like semantics of internal definitions that combine a defining computation and an application will be presented. By specialisation for a given defining computation one can derive inference rules for applications of defined operations.  
\\[2ex]
With semantics of that kind for every operation, each application identifies an axiom in a logic defined by the programming language, so a language can be considered a theory.
\end{minipage}
\end{center}

\section{\hhref[ref:contents]{Introduction}}

The Curry-Howard correspondence relates programming languages and logic.  Essentially it reflects similarity between propositions and proofs of a formal logic and types and programs of a programming language.  W.A.Howard is the author of the seminal paper on the correspondence.  
Various researchers have investigated its implications for programming. Transformer semantics is just for a type of language design based on notions that relates to \emph{weak existence} described by Howard separately in the same paper.

According to the Curry-Howard correspondence, the notions of programs correspond to proofs, and a program proves its type according to the typing rules of the programming language. So a powerful notion of types is needed for this to be of interest.  Per Martin Löf5~\cite{Martin-Lof:85} 
has advocated a system with such a type system.

Dijkstra's predicate transformers can be molded as a logic, but the predicates in such a system do not correspond to types, but programs correspond to proofs, if their applications are seen as recordings of proof steps.  Dijkstra's language statement constructs have the trivial type \emph{void} but in a language with a richer type system, typing rules would imply that programs seen as proofs would also prove the type of its result.

A formal system which includes expressions of the form \ensuremath{\TE{P}{R\sub\theta}} will be presented. These are mathematical expressions that combine the following parts: a program text \texttt{P}, a mathematical expression \ensuremath{R\sub\theta}, and an operator \TE{P}{} which might be compared to the integration operator 
\ensuremath{\int\_\_ \mbox{ \tt d}\theta}.
The operator \TE{P}{} binds free occurrences of $\theta$ in the expression \ensuremath{R\sub\theta} (as in lambda-calculus, say).  The intuitive meaning of the abstraction variable is \emph{an expression for the value obtained by interpretation of \texttt{P}}.  \TE{P}{} is an automorphism on mathematical expressions, partly related to Dijkstra's predicate transformers. 

The intuitive reading of $\TE{P}{R\sub\theta}$ is that \emph{\texttt{P} proves that an expression $\theta$ exists such that $R\sub\theta$ `is of interest'}.  The objective with $\TE{P}{R\sub\theta}$ is similar to Dijkstra's, i.e.~to obtain a mathematical expressions that, in the state from which \texttt{P} should be interpreted, has the value of $R_\theta$ in the in the result state.
 
The verb `to prove' is transitive, but it is possible to conceive the noun `proof' independently of an object as required by the verb.  A proof can be understood as a pattern of references to individual rules of a formal logic. A programmer writes programs that can be seen as proofs, when an adequate language is used.  Programmers do not need to master the logic aspects, but provide proofs much like people may do in every-day arguments.  Such proofs may need be checked but, hopefully, may be automated some day. 

A benefit of programs as proofs is that not all programmers need to master the logic aspects of programs completely.  One may still acknowledge programmers work as proofs in a substantial sense.  Language designers should, of course, be more concerned with the rules for application of their operations, i.e.~their adequateness as axioms of a logic.

Another benefit might be the implicit enforcement of a programming discipline that is indirectly influenced by formal logic, much the same as discussions in daily life may benefit from rules justified by formal logic.

\begin{comment}

The programs-as-proofs aspect of the Curry-Howard correspondence is corroborated by this work.  The propositions-as-types aspect only to a lesser degree.  One can illustrate this with an expression outline
\[\TE[\theta:T]{P}R_{\theta}\]
making it clear that $T$ and $R_\theta$ have different r\^oles.  Although it might be tempting to have $T$ express the set of primes (or even subsets of primes) it seems simpler to keep the notions apart.  Not the least with regard to acceptance by programmers.

A traditional proof of a program is merely a proof check that the given proof is effective in a given context.  Failure might stem from an incomplete case analysis, and that might lead to useful feed-back to programmers.

Mathematicans rarely rely on fully formal proofs, but formally state their goals.  So perhaps one might say: programmers rarely rely on formally stated goals, but formally state their proofs as programs.

\end{comment}

\section{\hhref[ref:contents]{Transformation semantics}}

Dijkstra has presented a notion of \emph{predicate transformers} along with a simplified programming language, and thus illustrated an important relationship between language design and a logic for program proofs: \cite{Dijk76}.  However, transformation rules appear to arise as axioms from the intuition of the language designer.  Dijkstra's language does not support definitions in programs.  But they are needed in practice and corresponding transfomation rules likewise. 

Rules for definitions within programs, and even inside definitions will be essential also to find rules for application of defined entities.  Such definitions will be called \emph{internal definitions}. It will be possible to write them in familiar style, but conceptually they differ a bit: a definition is more like familiar (letrec- or) let-constructs, i.e.~it combines one part that constitutes a conventional definition with an application part where the defined entity can be applied.

The notion of internal definitions in an adequate programming language that admits \texttt{P} in \TE{P}{} should allow definition of object-like entities, i.e.~means to introduce `members' associated with an `object' of some `class'.  An essential aspect is that programmers do not chose names of members, but accept names determined by the `class'.

Such introduction of names can be generalised, and the term \emph{implicit name binding} will be used for it as a design principle.  Furthermore, `in-line definitions' should be mandatory, which can be done by admitting first-order types only.  Higher-order entities will be admitted in another way.  In other words: operation names may not be used as arguments and implicit name binding thus required for every `in-line definition'.

An internal definition can be compared to a class with the defined entity as its only member in the application part.  Both notions combine details of semantics with a notion of abstraction over its application. Simula 67, \cite{Birtwistle80}, provides the  \emph{inner} notion for
the `remaining part of the current block', like the scope of definitions in several other languages, e.g.~Pascal and C.  Simula's \emph{inner} notion will be replaced by a (named) parameter, and the idea is generalised so that implicit name binding may occur in several arguments.

An essential difference for programmers, between classes and internal definitions, is the required description of `parameters' in the latter.  Identification of an internal definition includes the parameter description, over which one cannot abstract.  
Parameter descriptions will be structured like those of Pascal 
\cite{Steensgaard-Madsen79}, but extended with type parameters.  Higher-order parameters are then described by the same syntax as used for internal definitions.  The term \emph{signature} will be used for  parameter descriptions.

\section{\hhref[ref:contents]{A combinator for internal definitions}}

Several languages support a simple construct
\begin{center}\em
\Let\ f(x)= Expr\sub1 \In\ Expr\sub2 
\end{center}
where the name $f$ can be used in \textit{Expr\sub2} for a function that maps $x$ to \textit{Expr\sub1}.  It can be explained reasonably well with lambda-notation by  
$(\lam f.\mbox{\em Expr\sub2})(\lam x.\mbox{\em Expr\sub1})$, i.e.~an expression that can be rephrased as follows:
\[(\lam B.\lam A.(A(B))) (\lam x.\mbox{\em Expr\sub1}) (\lam f.\mbox{\em Expr\sub2})\]
With no name occurring free, the expression $\lam B.\lam A.(A(B))$ is a \emph{combinator} and can be given a global name, e.g.~DEF:
\[\mbox{DEF\ } (\lam x.\mbox{\em Expr\sub1})\ (\lam f.\mbox{\em Expr\sub2})\]
which, except for a missing signature and the explicit introduction of names, forms an internal definition. A single argument combinator is appropriate for a \emph{global definition} in agreement with mathematics,  but not for internal definitions.  This and the need to structure programs for human readers justifies internal definitions as a separate notion.

Lambda-calculus prescribe rewriting rules for expressions in lambda-notation.  Here we just depend on the $\beta$-rule:
\[
(\lam x.\mbox{\em Expr\sub1})(\mbox{\em Expr\sub2})=
\Subst[x]{\mbox{\em Expr\sub2}}\mbox{\em Expr\sub1}
\]
where the right-hand side prescribes substitution of \embox{Expr\sub2}  for $x$ in \mbox{\em Expr\sub1}. However, a reservation is necessary to avoid unintended name binding, i.e.~a free occurrence of a name in \embox{Expr\sub2} unintentionally getting bound in the process.  A request for a name change in such situations is a typical way to state the reservation. Our use of the substitution symbol, $\triangleleft$, shall include this reservation.

% \vfill\newpage\noindent
The following simplified example illustrates DEF and use of the $\beta$-rule. 
However, it also illustrates that lambda-calculus is unfit to render the proceedings in a way that is easy for humans to follow.  The many lambda-abstraction is one essential cause.  Another is that equational reasoning here blends the contexts of the two arguments where they should be kept apart to help readers,

\begin{quote}\label{ex:lambda}
\begin{alltt}
DEF\em
   (\ensuremath{\lambda}\,D\ensuremath{\lambda}\,A.A(\ensuremath{\lambda}\,X.D(X*2))) 
   (\ensuremath{\lambda}\,r.r(\ensuremath{\lambda}\,x.(x*x))(\ensuremath{\lambda}\,f.(f(3)))) 
=  (\ensuremath{\lambda}\,r.r(\ensuremath{\lambda}\,x.(x*x))(\ensuremath{\lambda}\,f.(f(3))))
       (\ensuremath{\lambda}\,D.\ensuremath{\lambda}\,A.A(\ensuremath{\lambda}\,X.D(X*2)))
=      (\ensuremath{\lambda}\,D.\ensuremath{\lambda}\,A.A(\ensuremath{\lambda}\,X.D(X*2)))
        (\ensuremath{\lambda}\,x.(x*x))(\ensuremath{\lambda}\,f.(f(3)))
=             (\ensuremath{\lambda}\,f.(f(3)))(\ensuremath{\lambda}\,X.(\ensuremath{\lambda}\,x.(x*x))(X*2))
=                  (\ensuremath{\lambda}\,X.(\ensuremath{\lambda}\,x.(x*x))(X*2))(3)
=                      (\ensuremath{\lambda}\,x.(x*x))(3*2)
=                          (3*2)*(3*2)
=  36
\end{alltt}
\end{quote}
Notice that the character case of names in the two arguments of DEF differ: all capitalised in the first, none in the second.  It reflects a simple ping-pong-like game of control.  

A programming language will be introduced in which an internal definition for the above can be expressed as:
\begin{quote}
\begin{alltt}
DEF r OF T [D[x:int]:intl] [A[f[X:intl]:int]:T]: T
  \{ A\{D\{X*2\}\} \}
  \{ r\{x*x\}\{f\{3\}\} \};
\end{alltt}
\end{quote}
The text after DEF on the first line is a \emph{signature} and contains enough information to avoid explicit parametrisation (i.e.~the $\lambda$-s).  The third line illustrates use of the defined operation \texttt{r} which itself has a structure similar to an internal definition: its first argument defines a function that is required to derive the function \texttt{f} used in the second argument.

From a signature one can automatically derive a rule for application of the derived operation, i.e.~\texttt{r} above:
\begin{equation}\label{eq:example1}% \small
\Rule[{\presume{
      \Rule[{\presume[]{
        \Rule{x}{\TE{\scope[D]{x}}}
        }}]{D\{\scope[D]{x}\}}{\TE{\scope{D}}}\\[1ex]
      \Rule[{\presume[]{
        \Rule[{\presume[]{
          \Rule{X}{\TE{\scope{X}}}
          }}]{f\{\scope{X}\}}{\TE{\scope[D]{f}}}
        }}]{A\{\scope[D]{f}\}}{\TE{\scope{A}}}
      }}]{r\{\scope{D}\}\{\scope{A}\}}{\TE{\scope[D]{r}}}
\end{equation}
where $\scope[D]{r}$ is the first argument of the internal definition.
Before we see how such a rule helps to derive the result value we need to explain the structure.

Rules have the form of equations or 
\ensuremath{\embox{conclusion}\dashv\embox{presumptions}} and are displayed as:
\[\Rule[{\presume[]{\itbox{presumption}_1\\\itbox{presumption}_2\\...}}]
{Left}{\TE{Right}}
\]
Each presumption can be a rule. Presumptions do hold by assumption.

\begin{Box}[t]\small
\Caption{axioms}{Axiom schemes}% \label{box:axioms}
\begin{enumerate}
\item DEF-scheme\hfill\\
\ensuremath{
    \Rule[{\presume[]{
      \defmac{f[P\sub{i}...]\sub{i=1..N\sub{f}}}
      {\scope[D]{},\scope{}}
    }}
    ]{DEF f...\,\{\scope[D]{f}\}\{\scope{f}\}}
    {\TE{\scope{f}}}
}
\item Annihilate\hfill\\
\ensuremath{\Rule{\{\}}{\Subst{\bullet}}}\hfill i.e.~error if $\theta$ occurs free in the (omitted) operand
\item Constant\hfill\\
\ensuremath{\Rule{C\,}{\Subst{C\,}}}\hfill with \texttt{C} a constant
\item Subsumed name translation (i.e.~used only without an explicit alternative)\hfill\\
\ensuremath{\Rule{X\,}{\Subst{X\,}}}\hfill NB: different fonts for X
\item Sequential computation\hfill\\
\Rule{\scope{1};\scope{2}}{\TE{\scope{1}}{\TE{\scope{2}}}}
\item Infix operator symbols (with \texttt{\#} varying over operator symbols)\hfill\\
\ensuremath{\TE{\scope{1}\#\,\scope{2}}=
\TE[\tau\sub1]{\scope{1}}\TE[\tau\sub2]{\scope{2}}
\Subst{(\tau\sub1\otimes\tau\sub2})}
\hfill where $\otimes$ identifie the meaning of \texttt{\#}
\end{enumerate}
\end{Box}

\section{\hhref[ref:contents]{Formalisation of copy-rule semantics}}

Semantics of internal definitions is formalised by a scheme of transformation rules that essentially formalises the copy-rule semantics of Algol 60 \cite{Naur:1963}.  The formalisation will be called the \emph{DEF-scheme} for brevity and it constitutes the only complex transformation rule that serves as an axiom.  Copy-rule semantics for Algol 60 has been described intuitively in a few words, but its formalisation is complex.

A version that covers only call-by-name parameters and with optional result types disregarded, forms an introduction to the general scheme.
\begin{equation}\label{eq:rule}
    \Rule[\presume{
      \defmac{f[P\sub{i}...]\sub{i=1..N\sub{f}}}
      {\scope[D]{},\scope{}}
    }
    ]{DEF f...\,\{\scope[D]{f}\}\{\scope{f}\}}
    {\TE{\scope{f}}}
\end{equation}
Symbol %
\ensuremath{\stackrel{\mathrm{\scriptsize def}}{=}} identifies a macro that combines a signature and a switch for control status, 
\verb|{|$D,E$\verb|}|, which indicates alternation between definition- and application-contexts.  This scheme is the interesting axiom of a logic of transformations. Box~\ref{box:axioms}~presents all of them.

The rules allow side-effects and imply call-by-name.  Call-by-value can be covered as expressed for application of an operation with signature
\begin{center}
\verb|with OF T,W (X:T) [Body(__:T):W] : W|
\end{center}
and semantics
\[
\Rule[{\presume{
      \Rule{Body\{\scope[D]{\_\_}\}}{\TE[\eta_1]{\scope[D]{\_\_}}\Subst[\_\_]{\eta_1}\TE{\scope{Body}}}
      }}]
{with\{\scope{X}\}\{\scope{Body}\}}{\TE[\eta_1]{\scope{X}}\Subst[X]{\eta_1}\TE{\scope[D]{with}}}
\]
Parentheses can be used in signatures to request call-by-value.  The example illustrates the obvious: that the axiom schemes above do not suffice, and that substitution is not not a trivial operation.

The form of internal definitions can be described with EBNF as
\begin{quote}
``\texttt{DEF}'' \emph{signature}  ``\verb|{|'' \scope[D]{\_} ``\verb|}|'' 
``\verb|{|''
\scope{\_} ``\verb|}|''\\
\mbox{\rm where } \emph{signature} =  
\emph{name} \{ ``\verb|[|'' \emph{signature} ``\verb|]|'' \}\sub{1..N}
\mbox{\rm [ {: \em type} ]}
\mbox{\rm and $N\geq0$}
\end{quote}
Signatures will be described in greater details later.

The DEF-scheme in Equation~\ref{eq:rule} tells that a definition of \texttt{f} combines two arguments
\scope[D]{f} and \scope{f}, and that the entire construct as
one expression is given by \TE{\scope{f}} with a presumption about
applications of $f$.

Presumptions in the DEF-scheme tell that application of a name
in one context is combined with a defining argument in the other.
This combination represents the formalisation of the copy-rule, with
their environment represented by presumptions.  

The complication of the DEF-scheme is expressed by the macro \defmac{\_}{\_,\_}:
\begin{equation}\label{eq:def}
\begin{array}[t]{@{}l}
  \defmac{
    f\sigs[i=1..N\sub{f}]
    { P\sub{i} \sigs[j=1..n\sub{i}]{g\sub{i,j}...} }}{\scope[D]{},\scope[E]{}}
  \\
  \hspace{0.5cm}\equiv
  \Rule[{
      \presume{
        \defmac{
          P\sub{i}
          % (s\sub{j})\sub{j=1..m\sub{P\sub{i}}}
          \sigs[j=1..n\sub{i}]{g\sub{i,j}...}}
        {\scope[E]{},\scope[D]{}}
      } \subox{i=1..N\sub{f}}
    }]
    { f\args[i=1..N\sub{f}]{\scope{P\subox{i}}} }
    { \TE{\scope[D]{f}}
    }% \hspace{0.25\textwidth}
  N\sub{f}\geq0
\end{array}
\end{equation}
A tool can generate DEF-schemes from signatures. Even from more complex signatures that may contain call-by-value parameters, type expressions, and (possibly overloaded) signatures of operator symbols.

A transformation rule for internal definitions is fairly complex.  A one-step expansion of the def-macro in the presumption of Equation~\ref{eq:def} may bring some relief:

\begin{equation}\label{eq:def2}
\begin{array}[t]{@{}l}
  \defmac{
    f\sigs[i=1..N\sub{f}]
    { P\sub{i} \sigs[j=1..n\sub{i}]{g\sub{i,j}...} }}{\scope[D]{},\scope[E]{}}
  \\
  \hspace{0.5cm}\equiv
  \Rule[{
     \presume{
         \Rule[{\presume{\defmac{g\sub{i,j}...}{\scope[D]{},\scope[E]{}}}\subox{j=1..n\sub{i}}}]
         {P\sub{i} \args[j=1..n\sub{i}]{\scope[D]{g\sub{i,j}}}}{\TE{\scope{P\sub{i}}}}
      } \subox{i=1..N\sub{f}}
    }]
    { f\args[i=1..N\sub{f}]{\scope{P\subox{i}}} }
    { \TE{\scope[D]{f}} }
  N\sub{f}\geq0
\end{array}
\end{equation}
Equation~\ref{eq:def2} expresses a presumption for application of DEF, which is a rule for applications of $f$. That rule is stated in terms of the unknown $\scope[D]{f}$, the first argument of DEF, the presumptions of which are similar rules for application of operations $P\sub{i}$. Applications of $P\sub{i}$ is stated in terms of unknowns $\scope{P\sub{i}}$ (i.e.~arguments of $f$).  The presumption of $\scope{P\sub{i}}$ are rules for applications of operations $g\sub{j}$ (i.e.~implicitly introduced names) in terms of unknowns $\scope[D]{g\sub{j}}$.  

Eventually a rule reduces to a pattern of substitutions corresponding to:
\begin{equation}\label{eq:T}
\Subst[f]{\Seq[.]{\lam P}{N\sub{f}}.\scope[D]{f}}\scope{f}
\hspace{2em}
\Subst[P\sub{i}]{\Seq[.]{\lam g}{n\sub{i}}.\scope{P\sub{i}}}\scope[D]{f}
\hspace{1.5em}
\end{equation}
The objective has been to obtain rules that do not use 
lambda-abstractions, but rather point-wise use of functions and implicitly introduced names.  Reservations about unintended name bindings associated with $\triangleleft$ apply for $\dashv$ also.

%
%
%
%

% \vfill\newpage\newpage
An important issue related to Equation~\ref{eq:def} is whether some language may satisfy the property.  To see that, we characterise a language as follows
\begin{itemize}
\item
Every operation must have a signature
\item
Application expression of the language has the form
\ensuremath{f\{\scope{1}\}\{\scope{2}\}...\{\scope{N\sub{f}}\}}
when the signature of $f$ is 

\texttt{
\begin{tabular}{ll}
f & [P\sub1 [g\sub{1,1} ...] [g\sub{1,2} ...] ... [g\sub{1,n1} ...]]\\
  & [P\sub2 [g\sub{2,1} ...] [g\sub{2,2} ...] ... [g\sub{2,n2} ...]]\\
  & ...\\
  & [P\sub{N\sub{f}}  [g\sub{N\sub{f},1} ...] [g\sub{N\sub{f},2} ...] ... [g\sub{N\sub{f},nN\sub{f}} ...] ] 
\end{tabular} }
\\[1ex]
where $N\sub{f}\geq0$ and $n\sub{k}\geq0$
\item
Each application expression is interpreted as a lambda-expression
\[{\cal T}(f\{\scope{1}\}\{\scope{2}\}...\{\scope{N\sub{f}}\}) = f(\underline{\lam g\sub1}.{\cal T}(\scope{1}))
(\underline{\lam g\sub2}.{\cal T}(\scope{2}))
...(\underline{\lam g\sub{N\sub{f}}}.{\cal T}(\scope{N\sub{f}}))\]
where \ensuremath{\underline{\lam g\sub{k}}} stands for
\ensuremath{\lam g\sub{k,1}.\lam g\sub{k,2}. ... \lam g\sub{k,nk}}

$\cal T$ introduces lambda-abstractions for implicitly introduced names and in agreement with the bindings expressed in Equation~\ref{eq:T}
\item
Each internal definition
\ensuremath{\ttbox{DEF}\ \ttbox{\em signature}\ \{\tt\scope[D]{}\} \{\tt\scope{}\}}
is an application of combinator DEF equipped with signature 
\begin{center}
\texttt{DEF\sub{\embox{signature}} [\embox{signature}] [App[\embox{signature}] ]}
\end{center}
\end{itemize}
Let this language be called $\cal L$.  The terms \emph{Howard~languages} and \emph{Howard~programs} will be used for languages characterised like $\cal L$ and their programs, respectively.

\begin{prop}
Language $\cal L$ satisfies
Equation~\ref{eq:rule}-\ref{eq:def2}, if adequate means ensure that names are distinct.
\end{prop}
\begin{proof}
The interpretation $\cal T$ implies that $\cal L$ is interpreted as a subset of lambda-expressions that satisfies Equations~\ref{eq:rule}-\ref{eq:def2} according to the correspondence stated in Equation~\ref{eq:T}.
\end{proof}
For a given signature, transformation rules for applications may be derived --- literally in case an internal definition actually exists and virtually in case a definition exists only in terms of some idealised Howard~language.  

Derivation is by \emph{specialisation}, which is similar to mathematical projection of a function of two variables by fixing the value of one.  An alternative term in programming language research is \emph{partial evaluation} and it caries over to logic in agreement with the Curry-Howard correspondence \cite{Jones:93:PartialEvaluation}.  The advantage of this approach is that programming details of some internal definition get eliminated by specialisation.  Furthermore it gives credit to the use of prototypes in system development.

It all means that every operation identifies a transformation rule for its application, and that program composition corresponds to combined use of inference rules.  In this sense a program is a proof, with the only addition that recursion in a definition is an appeal to proof by induction.

DEF-schemes can be considered key axioms of a logic with proofs encoded as Howard programs.  Other axioms, all briefly described in
% \Bref{axioms}
Box~\ref{box:axioms} 
are rather trivial.

%%%%%%%%%%%%%

% \addtolength{\textheight}{2\baselineskip}
\section{Equational reasoning with presumptions \label{sec:example1}}

\providecommand{\sep}{&}

A \emph{DEF-rule} is an instance of a given signature's DEF-scheme.    
Equation~\ref{eq:example1} illustrates such a DEF-rule, which is repeated here for your convenience:
\[
\Rule[{\presume[]{
      \Rule[{\presume[]{
        \Rule{x}{\TE{\scope[D]{x}}}
        }}]
{D\{\scope[D]{x}\}}{\TE{\scope{D}}}\\[1ex]
      \Rule[{\presume[]{
        \Rule[{\presume{
          \Rule{X}{\TE{\scope{X}}}
          }}]
{f\{\scope{X}\}}{\TE{\scope[D]{f}}}
        }}]
{A\{\scope[D]{f}\}}{\TE{\scope{A}}}
      }}]
{r\{\scope{D}\}\{\scope{A}\}}{\TE{\scope[D]{r}}}
\hspace*{2em}\mbox{(\ref{eq:example1})}
\]
Instantiation with {\tt\scope{D}{}=x*x, \scope{A}=f\{3\}} and 
{\tt \scope[D]{r}=A\{D\{X*2\}\}}
we obtain:
\[
\Rule[\presume{
      \Rule[{\presume[]{
        \Rule{x}{\TE{\scope[D]{x}}}
        }}]{D\{\scope[D]{x}\}}{\TE{x*x}}\\[1.5ex]
      \Rule[{\presume[]{
        \Rule[{\presume{
          \Rule{X}{\TE{\scope{X}}}
          }}]{f\{\scope{X}\}}{\TE{\scope[D]{f}}}
        }}]
{A\{\scope[D]{f}\}}{\TE{f\{3\}}}
      }]
{r\{x*x\}\{f\{3\}\}}   
{\TE{A\{D\{X*2\}\}}}
\]
Presumptions are unaffected by this initial step, but now we need to
instantiate the second with \texttt{\scope[D]{f}=D\{X*2\}} to get a rule by which the right-hand side of the equation can be transformed:

\[
\Rule[\presume{
      \Rule[{\presume[]{
        \Rule{x}{\TE{\scope[D]{x}}}
        }}]{D\{\scope[D]{x}\}}{\TE{x*x}}\\[1.5ex]
      \Rule[{\presume[]{
        \Rule[{\presume{
          \Rule{X}{\TE{\scope{X}}}
          }}]{f\{\scope{X}\}}{\TE{\scope[D]{f}}}
        }}]{A\{\scope[D]{f}\}}{\TE{f\{3\}}}\\
      \Rule[{\presume[]{
        \Rule[{\presume{
          \Rule{X}{\TE{\scope{X}}}
          }}]{f\{\scope{X}\}}{\TE{\scope[D]{f}}}
        }}]{A\{D\{X*2\}\}}{\TE{f\{3\}}}
      }]
{r\{x*x\}\{f\{3\}\}}   
{\TE{A\{D\{X*2\}\}}}
\]
The presumptions now consist of the two from the general rule and 
one instantiated from one of these.  We shall for brevity memorise the former to be instantiated when needed.  However, when an instantiated presumption is used, its presumptions get introduced.

\[
\begin{array}{@{=\ \ }l@{}l}
\multicolumn{1}{@{}l}{\TE{r\{x*x\}\{f\{3\}\}}}\\
\TE{f\{3\}}\presume[\sep]{
      ...\\
      \Rule[{\presume[]{
          \Rule{X}{\TE{\scope{X}}}
          }}]{f\{\scope{X}\}}{\TE{D\{X*2\}}}
      }\\[2ex]
\TE{D\{X*2\}}\presume[\sep]{
      \Rule[{\presume[]{
        \Rule{x}{\TE{\scope[D]{x}}}
        }}]{D\{\scope[D]{x}\}}{\TE{x*x}}\\
      ...\\
      \Rule{X}{\TE{3}}
      }\\[2ex]
\TE{x*x}\presume[\sep]{
      ...\\
      \Rule{X}{\TE{3}}\\[1ex]
      \Rule{x}{\TE{X*2}}
      }\\[2ex]
\TE[\tau_1]{X*2}\TE[\tau_2]{X*2}\Subst{(\tau_1\cdot\tau_2)}\ \presume[\sep]{
      ...\\
      \Rule{X}{\TE{3}}
      }
\end{array}
\]
Continued transformation by rules for infix operators trivially leads to the value 36 in agreement with the lambda-calculus example (page~\ref{ex:lambda}).

% \clearpage %% just testing rather than trying to squize text on p.12

\section{\hhref[ref:contents]{Language design and transformation semantics}}

Dijkstra has introduced \emph{predicate transformers} as semantics of a programming language, striving to establish programming as a discipline of mathematics.
Semantics had before that mostly been presented as mathematical models, with the notion of \emph{state} as a stumble point for mathematicians.  The important advantage of using transformations is elimination of state changes, rather than explaining a process in terms of a sequence of states.

Dijkstra's semantics did not include definitions.  Internal definitions as introduced above depend on context and their semantics is given by DEF-schemes.  So Howard languages illustrate that transformation semantics can be expressed for languages more general than Dijkstra's.

A tool exists to implement Howard languages with support for a fixed, common syntax and the following concepts:
\begin{itemize}
\item
internal definition with signatures (of a fairly general notion of operations)
\item
operation applications including infix notation with operator symbols
\item
usual notation for sequential composition of computations
\item
type inference with types considered sets (in the mathematical sense)
\end{itemize}
% \addtolength{\textheight}{-2\baselineskip}
Essentially there are so few predefined operations that the core can hardly be classified as a programming language.  The tool helps introduce predefined operations from a given signature.

Support for one family of languages can be considered both negative and positive from a language designer's point of view. Language implementers are restricted syntactically but provided with strong support for type checking and modularity.  The tool represents a conservative choice of syntax and allows conveniences where a strict syntactic reflection of concepts is undesirable (e.g.~'declaration' as a shorthand for an application).  

% signatures as semantics of internal definitions
\subsection{\hhref[ref:contents]{Semantics of signatures}}

\begin{ebook}
\renewcommand{\topfraction}{0.8}
\renewcommand{\textfraction}{0.2}
\end{ebook}
The term `signature' is related to similar concepts in other programming languages, and some might easily and reasonably consider them essentially identical.  Some consider it a simplified version of a concept not yet clarified and blame it for not covering semantics of the operation with a given signature. 

Before going on, recall that internal definitions are similar to classes while an item being defined is similar to a member.  Probably we agree that the semantics of classes and members should not be confused, so likewise we need to distinguish between semantics of an internal definition and the item being defined.  So: 
\begin{quote}
\emph{A signature encodes semantics of internal definitions+}
\end{quote}
Concretely it means that a signature can be translated into a transformation rule for possible instances of
internal definitions with the given signature.

\subsection{\hhref[ref:contents]{Signatures and command syntax}}

Signatures determine the syntax of named commands as given below in EBFN: a pair of braces contains patterns that can be iterated (or omitted), a pair of bracket contains patterns that can be omitted.  Actual symbols appear in typed font between ``\,- and ''-characters.  
\begin{description}
\item[Signatures identified by names]~\hfill\\
    Signature  =\ Name {[} TypeInf {]} 
    \{ ``\texttt{\tt[}'' Signature ``\texttt{\tt]}'' \} 
    { [ ``\texttt{:}''} Type {]} $|$\\
    TypeInf\hspace{.8em} = {``\texttt{OF}''} [ Numeral ] Name 
    \{ ``\texttt{,}'' [ Numeral ] Name \}
\item[Application syntax]~\hfill\\\hspace*{1em}
\begin{tabular}{@{\hspace*{-1em}}l@{~}l}
    Expr & =\ Name \{ [ Label [ ``\texttt{:}'' ] ] Arg \} $|$
    ``\texttt{ DEF}'' Signature Arg Arg \\
    Arg    & =\ ``\verb|{|'' Expr \{ ``\texttt{;}'' Expr \}
                  ``\verb|}|''
 \end{tabular}
\end{description} 
The number of required arguments in an application is equal to the
length of the list of signatures following the name in the operation's signature.  The syntax above is simplified to emphasise the correspondence between signatures and arguments. Box~\ref{box:syntax}
provides a more complete description.

\begin{Box}[t]\smallsf
\caption{Signatures and expressions\label{box:syntax}}
%\begin{quote}
  \begin{tabular}{ll}
    Signature &= BasicSignature $|$ OpSignature $|$ CbVSignature\\[1ex]
    
    BasicSignature &= Identification \{ ``\texttt['' Signature ``\texttt]'' \} [ ``\texttt:'' TypeExpression ] \\
    Identification &= Name [ ``\texttt{OF}'' TypeOperator \{ ``\texttt,'' TypeOperator \}] \\
    TypeOperator &= [ NaturalNumber ] Name \\
    TypeExpression &= \{ TypeExpression \} Name
\\[1ex]

    CbVSignature &= Identification ValueList \{ ``\texttt{\tt[}" Signature ``\texttt{\tt]}'' \}\\ %
                 &\hspace{2em}``\texttt:" TypeExpression \\ %
    ValueList &= ``\texttt{(}`` TypedSignature \{ ``\texttt,'' TypedSignature \} ``\texttt)''\\ %
    TypedSignature &= Identification [ ValueList ] ``\texttt:'' TypeExpression\\ %
          &\hspace{2em}$|$ OperatorSignature\\[1ex]
 
    OpSignature 
      &= OperatorId
         ``\texttt(`` TypeExpression ``\texttt,'' TypeExpression ``\texttt)''\\
          &\hspace{2em} ``\texttt:'' TypeExpression \\ 
    OperatorId
      &= OperatorSymbol [ ``\texttt{OF}'' TypeOperator \{``\texttt,'' TypeOperator\} ]%
      \\
    \end{tabular}
%\end{quote}

%\begin{quote}
  \begin{tabular}{lcl}
  Expression &=& E $|$ [ Qualifier ``\texttt{.}'' ] Name \{ Argument $|$ ArgList\}\\
  Qualifier   &=& LevelName\\
  Argument    &=& [ LevelName ] ``\verb|{|'' Expression \{``\texttt;'' Expression \} 
  [ ``\texttt;" ] ``\verb|}|''\\
  ArgList	  &=& [ LevelName ] ``\texttt{(}`` Expression \{ ``\texttt,'' Expression \} ``\texttt)''\\
  LevelName   &=& Name [ ``\texttt:'' ]\\
  E           &=& Application $\ |\ $ [ O ] [ E ] \{ [ E ] O E \} [ E ] [ O ] $\ |\ $ ``\texttt('' E ``\texttt)''\\
  O           &=& OperatorSymbol
  % Expression \{ Operator Expression \}
\end{tabular}\\
A \texttt{?}-symbol is assumed, if no other operator separates two expressions. 
%\end{quote}
\end{Box}

% {\samepage
\noindent
\emph{Convenience rules} prescribe some special notations as identical to expressions that adhere to the syntax:
\begin{description}
\item[Identity] is a braced, semicolon separated lists of expressions not being an argument of an application.  Identities are 
are interpreted as arguments of a \texttt{program} operation which is a polymorphic identity.
\item[Declaration] allows an application to be written with its last argument apparently missing, but present as `the remaining part of a context'.  This is often used to write an internal definition as in other languages, and similarly use common notation for instantiation of a class.
\item[Syntactic coercion] allows a level name to be used by itself, provided a name for a \emph{default member} is introduced in the argument it applies to.  The default member name is by convention \texttt{\_\_}.  Further, when alse a name \texttt{\_} is introduced, the default member name may be replaced by that when used as a call-by-value argument, or according to descriptions of operator symbols.
\item[A list expression] is a bracketed, comma separated lists of expressions.  Such ar interpreted as a \texttt{::} separated lists
of the expressions terminated by \ttbox{::nil}, with the operator symbol and \texttt{nil} being user defined.
\end{description}
% }
% \vfill\newpage\noindent
Language designers who want mutable \emph{variables} may provide an operation for their declaration with a signature like
\begin{quote}
\begin{verbatim}
    var OF w, rvalue [Scope OF lvalue
                        (__:lvalue)
                        [_:rvalue]
                        [:=(lvalue,rvalue) : rvalue]:w]:w
\end{verbatim}
\end{quote}
Operation \texttt{var} may be applied as in 
\verb|{ var x; ...; x:=x+1; ... }| 
which will be disambig\-uated as \verb|{ var x{ ...; x.__:=x._+1; ... } }|,
i.e.~by taking `the rest of the brace' as the missing argument in an
application written as a `declaration'

\subsection{\hhref[ref:contents]{Example: induction}}

An operation can be defined to match the structure of proofs by induction: \label{ex:induction}
\begin{verbatim}
    induction OF Problem, Result
       [Initial:Problem]
       [Break_down[__:Problem]
          [result[Sub:Problem]:Result]:Result]:Result
\end{verbatim}
and it might be useful in exercises to teach programming with easy-to-prove iterations, e.g.~to compute the product of elements in a list use
\begin{verbatim}
  induction{[3,5,7]} L: { split_list{L}{1}{hd*result{tl}} }
\end{verbatim}

\section{\hhref[ref:contents]{Specialisation}}

A well-known example, \texttt{twice}, can be internally defined and used as follows
\begin{quote}\small
\begin{verbatim}
{ DEF twice [F[x:int]:int] [Return[f[X:int]:int] :int]:int 
  { Return{F{F{X}}} };  
  twice{x*x};  # i.e. F is x -> x*x  with implicit x
  f{2}         # so f(2) is (2*2)*(2*2)
}
\end{verbatim}
\end{quote}
So, although only values can be returned, it is possible to provide access to a
function as a member, say `f', of a class-like operation.  The example uses the syntactic convenience rule for `declarations' that makes it equivalent to
\begin{quote}\small
\begin{verbatim}
{ DEF twice [F{x:int}:int] [Return[f{X:int}:int] :int]:int 
  { Return{F{F{X}}} }
  { twice{x*x}{f{2}} }
}
\end{verbatim}
\end{quote}
i.e.~the apparently missing argument is taken as `the rest of the brace'.  In this form there is no need for the outermost braces.

It is bad to restrict the result of the entire computation to \texttt{int}. but the
type systems allows the following which calls for type inference to bind the type name
\texttt{W} appropriately.
% \vfill\newpage
\begin{quote}\small
\begin{verbatim}
{ DEF twice OF W [F[x:int]:int] [Return[f[X:int]:int] :W]:W 
  { Return{F{F{X}}} };
  twice{x*x};  # i.e. F is x -> x*x  with implicit x
  {f{2}}       # so f{2} is 2-> {2*2}*{2*2}
}
\end{verbatim}
\end{quote}
in this case with the same result, but now \verb|f{2}| can be replaced by something like \verb|stdout << f{2} << nl| of type \texttt{Output}.

\texttt{INTERPRET} is the name of a special operation that interprets an expression in context, as for instance in a nested read-eval-print-loop
\begin{quote}\small
\begin{verbatim}
DEF twice OF W [F[x:int]:int] [Return[f[X:int]:int] :W]:W 
  { Return{F{F{X}}} }
  { twice{x*x}        # i.e. F is x -> x*x  with implicit x
    {loop{INTERPRET}} # expresses a read-eval-print-loop
  }
\end{verbatim}
\end{quote}
so that an input of \verb|f{2}| will return 16.

An operation, \texttt{defrule}, defined internal to the interpreter can be used to generate \LaTeX-text for inclusion in documentations, and in this case \verb|defrule{"twice"}| has been used to for the manuscript to include:
\begin{quote}
\begin{verbatim}
    twice OF W
       [F[x:int]:int]
       [Return[f[X:int]:int]]
\end{verbatim}

\ensuremath{\Rule[\presume{
      \Rule[{\presume[]{
        \Rule{x}{\TE{\scope[D]{x}}}
        }}]
{F\{\scope[D]{x}\}}{\TE{\scope{F}}}\\
      \Rule[{\presume[]{
        \Rule[\presume{
          \Rule{X}{\TE{\scope{X}}}
          }]
{f\{\scope{X}\}}{\TE{\scope[D]{f}}}
        }}]
{Return\{\scope[D]{f}\}}{\TE{\scope{Return}}}
      }]
{twice\{\scope{F}\}\{\scope{Return}\}}{\TE{\scope[D]{twice}}}}
\end{quote}

% \addtolength{\textheight}{2\baselineskip}
\subsection{\hhref[ref:contents]{\label{app:specialisation}Details: Specialising operation \texttt{twice}}}

The computational definition of \texttt{twice} is known and can be used to derive a rule for application of \texttt{twice}.  The idea is similar to the projection of a function of pairs of values by keeping one fixed.

Substitution of \verb|Return{F{F{X}}}| for \texttt{\scope[D]{twice}}
and the consequential \verb|F{F{X}}| for \texttt{\scope[D]{f}} gives
%%%
\begin{quote}
\ensuremath{
\Rule[\presume{
\begin{comment}
      \Rule[{\presume[]{
        \Rule{x}{\TE{\scope[D]{x}}}
        }}]{F\{\scope[D]{x}\}}{\TE{\scope{F}}}\\
      \Rule[{\presume[]{
        \Rule[\presume{
          \Rule{X}{\TE{\scope{X}}}
          }]{f\{\scope{X}\}}{\TE{\scope[D]{f}}}
        }}]{Return\{\scope[D]{f}\}}{\TE{\scope{Return}}}
\end{comment}
        ...\\
      \Rule[{\presume[]{
        \Rule[\presume{
          \Rule{X}{\TE{\scope{X}}}
          }]{f\{\scope{X}\}}{\TE{F\{F\{X\}\}}}
        }}]{Return\{F\{F\{X\}\}\}}{\TE{\scope{Return}}}
      }]{twice\{\scope{F}\}\{\scope{Return}\}}
        {\TE{Return\{F\{F\{X\}\}\}}}
}
\end{quote}
Recall that elipsis denote memorised presumptions.
We can instantiate the first presumption twice, once with \verb|F{X}|
and once with \texttt{X} substituted for \texttt{\scope[D]{X}}: 
\begin{quote}
\ensuremath{
\Rule[\presume{
\begin{comment}
      \Rule[{\presume[]{
        \Rule{x}{\TE{\scope[D]{x}}}
        }}]{F\{\scope[D]{x}\}}{\TE{\scope{F}}}\\
      \Rule[{\presume[]{
        \Rule[\presume{
          \Rule{X}{\TE{\scope{X}}}
          }]{f\{\scope{X}\}}{\TE{\scope[D]{f}}}
        }}]{Return\{\scope[D]{f}\}}{\TE{\scope{Return}}}
      q...\\
      \Rule[{\presume[]{
        \Rule[\presume{
          \Rule{X}{\TE{\scope{X}}}
          }]{f\{\scope{X}\}}{\TE{F\{F\{X\}\}}}
        }}]{Return\{F\{F\{X\}\}\}}{\TE{\scope{Return}}}
\end{comment}
...\\
      \Rule[{\presume[]{
        \Rule{x}{\TE{F\{X\}}}
        }}]{F\{F\{X\}\}}{\TE{\scope{F}}}\\
      \Rule[{\presume[]{
        \Rule{x}{\TE{X}}
        }}]{F\{X\}}{\TE{\scope{F}}}\\
      }]
{twice\{\scope{F}\}\{\scope{Return}\}}{\TE{Return\{F\{F\{X\}\}\}}}
}
\end{quote}
The two rules can now be combined to one as in
\begin{quote}
\ensuremath{
\Rule[\presume{
	  ...\\
      \Rule[{\presume[]{
        \Rule{x}{\TE{F\{X\}}}\\
        \Rule[{\presume[]{
          \Rule{x}{\TE{X}}
          }}]{F\{X\}}{\TE{\scope{F}}}
        }}]{F\{F\{X\}\}}{\TE{\scope{F}}}\\
      }]
{twice\{\scope{F}\}\{\scope{Return}\}}{\TE{Return\{F\{F\{X\}\}\}}}
}
\end{quote}
This rule can be simplified by use of the third presumption to
\begin{quote}
\ensuremath{
\Rule[\presume{
	  ...\\
      \Rule[{\presume[]{
        \Rule{x}{\TE{F\{X\}}}\\
        \Rule[{\presume[]{
          \Rule{x}{\TE{X}}
          }}]{F\{X\}}{\TE{\scope{F}}}
        }}]{F\{F\{X\}\}}{\TE{\scope{F}}}\\
      }]
{twice\{\scope{F}\}\{\scope{Return}\}}{\TE{\scope{Return}}}
}
\end{quote}
% \addtolength{\textheight}{-2\baselineskip}
The bottom two presumptions can be combined to eliminate 
\texttt{\TE{F\{F\{X\}\}}}, and at the same time drop the two presumptions at the top:
\begin{quote}
\ensuremath{
\Rule[\presume{
      \Rule[{\presume[]{
        \Rule{X}{\TE{\scope{X}}}\\
        \Rule{x}{\TE{F\{X\}}}\\
        \Rule[{\presume[]{
          \Rule{x}{\TE{X}}
          }}]{F\{X\}}{\TE{\scope{F}}}
          }}]{f\{\scope{X}\}}{\TE{\scope{F}}}\\
      }]
{twice\{\scope{F}\}\{\scope{Return}\}}{\TE{\scope{Return}}}
}
\end{quote}
Finally the presumptions of the presumption can be combined so that every explicit reference to the defining context gets eliminated:
\begin{quote}
\ensuremath{
\Rule[\presume{
      \Rule[{\presume[]{
        \Rule[{\presume[]{
          \Rule{x}{\TE{\scope{X}}}
          }}]{x}{\TE{\scope{F}}}
          }}]{f\{\scope{X}\}}{\TE{\scope{F}}}\\

      }]
{twice\{\scope{F}\}\{\scope{Return}\}}{\TE{\scope{Return}}}
}
\end{quote}
Applications of operation \texttt{f} in \texttt{\scope{Return}} is the
given by the rule
\begin{quote}
\ensuremath{
      \Rule[{\presume[]{
        \Rule[{\presume[]{
          \Rule{x}{\TE{\scope{X}}}
          }}]{x}{\TE{\scope{F}}}
          }}]{f\{\scope{X}\}}{\TE{\scope{F}}}
}
\end{quote}
So for any \texttt{int}-constant c we have
\[
\begin{array}{l}
\multicolumn{1}{@{}l}{\TE{f\{c\}}}\\
 = \TE{x*x}\presume[]{
        \Rule[{\presume[]{
          \Rule{x}{\TE{c}}
          }}]{x}{\TE{x*x}}
          }\\
 = \TE[\tau\sub1]{x}\TE[\tau\sub2]{x}\Subst{(\tau\sub1\times\tau\sub2)}
\presume[]{
        \Rule[{\presume[]{
          \Rule{x}{\TE{c}}
          }}]{x}{\TE{x*x}}
          }\\
= \TE[\tau\sub1]{x*x}\TE[\tau\sub2]{x*x}
\Subst{(\tau\sub1\times\tau\sub2)}\presume[]{\Rule{x}{\TE{c}}}
\\
= \TE[\mu\sub1]{x}\TE[\mu\sub2]{x}
\Subst[\tau\sub1]{(\mu\sub1\times\mu\sub2)}
\TE[\mu\sub5]{x}\TE[\mu\sub6]{x}
\Subst[\tau\sub2]{(\mu\sub3\times\mu\sub4)
}
\Subst{(\tau\sub1\times\tau\sub2)}\presume[\\\hspace*{1em}]{\Rule{x}{\TE{c}}}
\\
= \TE[\mu\sub1]{c}\TE[\mu\sub2]{c}
\Subst[\tau\sub1]{(\mu\sub1\times\mu\sub2)}
\TE[\mu\sub5]{c}\TE[\mu\sub6]{c}
\Subst[\tau\sub2]{(\mu\sub3\times\mu\sub4)
}
\Subst{(\tau\sub1\times\tau\sub2)}\\
= \Subst{(c\times c)\times(c\times c)}
\end{array}
\]
% \end{quote}
%%

\section{\hhref[ref:contents]{Sideeffects}}

While a program is being developed, programmers sometimes need to add
code to reflect progress (read: problematic behaviours).  Mostly it amounts to what is known as `side-effects' because it really illustrates that programs differ from mathematical expressions in kind.

An illustration from an application of operation \texttt{twice} with a read-eval-print-loop is
\begin{center}\small
\begin{verbatim}
.demo  1> {var t; t:=0; f{stdout << ~"Step " << t:=t+1 << nl; 2} };
Step 1
Step 2
Step 3
Step 4
16
\end{verbatim}
\end{center}
which reveals that the argument of operation \texttt{f} is `computed' four times to arrive at the result 16, as postulated earlier.

This is actually reflected in the rewriting to evaluate \verb|f{c}| in the previous section: the step prior to the last contains 
\TE[\mu\sub{k}]{c} for k=1-4.  Obviously this is a point where efficiency can become important, and hence makes a need for means to avoid such repetitions.  In other words: a need for call-by-value arguments.

Arguments for which no name is introduced implicitly are the points where the issue of such side-effects may arise.  The term \emph{call-by-name} is used when such an argument is computed whenever the interpreter refers to it.  A companion term is \emph{call-by-value} which implies that a single reference is used to obtain the argument's value, which is then saved for later use.

The complete syntax of signatures allows call-by-value to be requested, but notice that it is a programmer's responsibility to meet such requests.  The actual syntax is simply to allow parentheses instead of brackets  around a typed signature.  In the example all brackets could be replaced by corresponding parentheses.  Operation \texttt{with} evaluates its first argument as a value and passes it to its second argument. 

The following definition and its use illustrates how call-by-value can be achieved 
\begin{center}\small
\begin{verbatim}
 2>  DEF twice OF W [F(x:int):int] [Return[f(X:int):int] :W] : W
 3>   { Return{with(X) u; F{F{u}}} }
 4>   { twice{x*x}        # i.e. F is x -> x*x  with implicit x
 5>     {loop{INTERPRET}} # similar to loop-eval-print
 6>   };
. 1> {var t; t:=0; f{stdout << ~"Step " << t:=t+1 << nl; 2} };
Step 1
16
\end{verbatim}
\end{center}
More serious problems are that requests for call-by-value should be reflected in the semantics of an internal definition, and that means to express it should be found. The problems have been solved, as illustrated in the following subsection.

\subsection{\hhref[ref:contents]{Call-by-value vs.~call-by-name}}

\begin{ebook}
\addtolength{\textheight}{2\baselineskip}
\end{ebook}

Operation \texttt{defrule("twice")} generates the manuscript for the following display after the signature has been changed:
\begin{quote}
%\begin{Howard}\itt{twice}
\begin{verbatim}
    twice OF W
       [F(x:int):int]
       [Return[f(X:int):int]:W]:W
\end{verbatim}

\ensuremath{\Rule[\presume{
      \Rule{F\{\scope[D]{x}\}}{\TE[\eta_1]{\scope[D]{x}}\Subst[x]{\eta_1}\TE{\scope{F}}}\\
      \Rule[{\presume[]{
        \Rule{f\{\scope{X}\}}{\TE[\eta_1]{\scope{X}}\Subst[X]{\eta_1}\TE{\scope[D]{f}}}
        }}]
{Return\{\scope[D]{f}\}}{\TE{\scope{Return}}}
      }]
{twice\{\scope{F}\}\{\scope{Return}\}}{\TE{\scope[D]{twice}}}}
%\end{Howard}
\end{quote}
First note the use of parentheses instead of brackets in the signature.  Second note that \texttt{\scope[D]{x}} on the right-hand side in the conclusion of the first presumption tells that the mathematical expression bound to the $\eta\sub1$ is substituted for $x$ once and that such an expression cannot contain side-effects.
Of course this implies that the previous description of the
\defmac{\_\_}{\_\_} is incomplete and the complete version has been implemented in the tool.

% \vfill\newpage
It makes a nice exercise for readers to specialise the above rule for definition of \texttt{twice} to a rule for applications of \texttt{twice}  until the following point is reached:
\begin{quote}
\ensuremath{\Rule[\presume{
      \Rule{f\{\scope{X}\}}{\TE[\eta_1]{\scope{X}}\Subst[X]{\eta_1}\TE{F\{F\{X\}\}}}\\
      \Rule{F\{F\{X\}\}}{\TE[\eta_1]{F\{X\}}\Subst[x]{\eta_1}\TE{x*x}}\\
      \Rule{F\{X\}}{\TE[\eta_1]{X}\Subst[x]{\eta_1}\TE{x*x}}\\      }]
{twice\{x*x\}\{f\{c\}\}}{\TE{f\{c\}}}}
\end{quote}
The next step is to use the presumptions by equational reasoning:
\[
\Rule{f\{c\}}{\TE[\eta_1]{c}\Subst[X]{\eta_1}\TE{F\{F\{X\}\}}}\\
=\TE[\eta_1]{c}\Subst[X]{\eta_1}\TE[\eta_1]{F\{X\}}\Subst[x]{\eta_1}\TE{x*x}
\]
Now we face a situation where substitution is tricky.  The $\eta\sub1$ in \ensuremath{\Subst[X]{\eta\sub1}} is a free occurrence that will become bound if the substitution is done thoughtlessly, since every occurrence of $\eta\sub1$ is bound by \ensuremath{\TE[\eta\sub1]{F\{X\}}}.  The problem is overcome by renaming (i.e.~the rule called $\alpha$-conversion 
in the lambda-calculus).  Thus we may proceed
\[
\begin{array}{l@{\ =\ }l}
\cdots
& \TE[\eta\sub2]{c}\Subst[X]{\ensuremath{\eta\sub2}}
\TE[\eta\sub1]{F\{X\}}\Subst[x]{\ensuremath{\eta\sub1}}\TE{x*x}\\
& \TE[\eta\sub2]{c}\Subst[X]{\ensuremath{\eta\sub2}}
\TE[\eta\sub3]{X}\Subst[x]{\ensuremath{\eta\sub3}}\TE[\eta\sub1]{x*x}
\Subst[x]{\ensuremath{\eta\sub1}}\TE{x*x}\\
& \TE[\eta\sub2]{c}\Subst[X]{\ensuremath{\eta\sub2}}
\TE[\eta\sub3]{X}\Subst[x]{\ensuremath{\eta\sub3}}\Subst[\eta\sub1]{(x\times x)}
\Subst[x]{\ensuremath{\eta\sub1}}\Subst{(x\times x)}\\
& \TE[\eta\sub2]{c}\Subst[X]{\ensuremath{\eta\sub2}}\TE[\eta\sub3]{X}
\Subst[x]{\ensuremath{\eta\sub3}}\Subst{((x\times x)\times(x\times x))}\\
& \TE[\eta\sub2]{c}\Subst[X]{\ensuremath{\eta\sub2}}
\Subst{((X\times X)\times(X\times X))}\\
& \Subst{((c\times c)\times(c\times c))}
\end{array}
\]
Enough details have been presented so that a careful reading will show where renaming is required to avoid unintended name bindings.

\begin{comment}
\addtolength{\textheight}{-2\baselineskip}
\end{comment}

\section{\hhref[ref:contents]{Verification}}
Program verification implies application of the transformation rules along with some expression that may be a predicate.  Rules can be identical to familiar rules introduced by others, e.g.~Hoare's rule for assignments.  A rule for use of operation \texttt{induction}, cf.~page~\pageref{ex:induction} matches proof by mathematical induction.  Side-effects are essentially described like assignments to hidden variables.

Two examples follow to illustrate (a) elimination of program variables and (b) informal verification based on equivalence of program fragments.\vspace{2ex}

\begin{description}% \samepage
\item[Elimination]~\hfill\\
Hoare's rule for assignments is: \ensuremath{\TE{x:=\scope{}}=\TE{\scope{}}\Subst[x]{\theta}}.  
Right-to-left progress implies that an expression using variable \texttt{x} in the state just after its initialisation can be quite complex but have the form ${\cal E}_x$.  Initialisation implies that $x$ gets substituted by some expression, often a constant, so that $x$ gets eliminated from the resulting mathematical expression: 
$\Subst[x]{e}{\cal E}_x$.
\item[Induction]~\hfill\\
Mathematical induction proves
\[
(\TE{induction(n)\,i:\{if(i=0,0(2*i-1)+result(i-1))\}}\theta)=\Sigma_{i=0}^n(2i-1)
\]
\begin{paper}\newpage\end{paper}
\item[Equivalence]~\hfill\\
A slight variation of the previous example is
\[\TE{var \,x\{x:=0; induction(n)\,i:\{if(i=0,x)\{result(i-1);\,x:=2*i-1\}\};\,y:=x\}}{\cal P}
\]
which transforms to
\begin{equation}\label{eq:induction}
\Subst[x]{0}\TE{induction(n)\,i:\{if(i=0,x)\{result(i-1);\,x:=x+2*i-1\}\}}\TE{x}\Subst[y]{\theta}{\cal P}
\end{equation}
With the presumption as induction hypothesis we can prove
\[
\Rule[{
\presume{
  % \Rule[{\presume{k=0}}]{result(k)}{x}\\
  \Rule[{\presume[]{i<k}}]{result(i)}{\Subst{(x+\Sigma_{j=0}^k(2j-1))}}  
}}]{result(k)}{\TE{induction(k)\,i:\{if(i=0,x)\{result(i-1);\,x:=x+2*i-1\}\}}}
\]
and conclude: 
\[\forall n.\TE{induction(n)\,i:\{if(i=0,x)\{result(i-1);\,x:=x+2*i-1\}\}}\theta=x+\Sigma_{i=0}^n(2i-1)
\]
The entire expression~\ref{eq:induction} then becomes
\[
\Subst[x]{0}\Subst{(x+\Sigma_{i=0}^n(2i-1))}\Subst[x]{\theta}\Subst[y]{\theta}{\cal P}
\]
which finally can be reduced to
\[
\Subst{\Sigma_{i=0}^n(2i-1)}\Subst[x]{\theta}\Subst[y]{\theta}{\cal P}
\]
so $x$ is eliminated but its final value may survive elsewhere in the context.

A more efficient computation results, if application of \texttt{induction} is replaced by an equivalent application of \texttt{while}.  The rule for the former is the familiar and obvious, and the path followed here might be useful in other situations.
\end{description}
Detailed inference rules have not been presented, for brevity and because this should not be considered an advocacy for specific operations.  But hopefully the examples are illustrative anyway.

\subsection{\hhref[ref:contents]{Partial and total correctness}}

With rules as presented, programs as proofs can justify partial correctness only, i.e. proofs presume program termination.  So a separate proof of termination is required.

With generated transformation rules, termination is ensured only if every program step terminates, so one may have to add manually presumptions to ensure termination.  Sometimes that requires knowledge about a specific application domain.

Signatures may exist for operations that belong to a specific field of applications.  An example is presented later and concerns unix-style process control (see page~\pageref{unix}).  Termination properties with the chosen `members' are complex and it may be impossible to reduce them to the usually known techniques for proofs of termination.  A better design may, perhaps, encapsulate the problem.

Howard languages are adequate as domain specific languages, and the issue of termination has to be delegated to specialist from the application field.  How to help a specialist design operations with appropriate termination properties is an open question. 

% modularity
\begin{comment}
\vfill\newpage
\end{comment}

\section{\hhref[ref:contents]{Components and their composition}}

The notion of a component in this context is delightfully easy and general: 
\begin{center}
\emph{A component is a named operation with a signature}
\end{center}
It means that the notion of a named operation is very broad, covering simple functions, structured statements, class-like entities, members that are operations, and generalised classes with several arguments each with own members.

Composition is the same for all kinds of operations and corresponds
to function composition in the simplest case and statement composition otherwise.  \emph{Implicit name introduction} and the simple notion of types as sets have been essential for this simplification.  Sets of functions are thus not admitted.

Transformation rules for internal definitions are expressed in terms of unknown program fragments.  Each application implies a binding of unknowns to actual fragments and thus makes specialisation possible.  Transformation and evaluations work in opposite directions: the former right-to-left the latter left-to-right.

Design of components can be a complex task.  But a strict regime as represented by the requirement of a signature (as presented) for every component tends to help in design.

One thing is components of a single program, another components of collections of programs.  Experiences tells that a collections of programs may be connected by channels, e.g.~POSIX pipes.  Texts sent over a channel can be interpreted as program expressions in a context at the receiving side.  One important predefined operation supports this behaviour in an actual interpreter.  

% programming language design elements: operations, notation, semantics, 
%  and implementation

% \section{\hhref[ref:contents]{Illustrative signatures}}
% \vfill\newpage

% \vfill\newpage
\subsection{\hhref[ref:contents]{UNIX-based control operations}}

An operation \texttt{unix} has been implemented with a signature shown in Box~\ref{box:UNIX}.
\begin{Box}[t]\caption{Signature of operation \texttt{unix}\label{box:UNIX}}
\label{unix}
\begin{verbatim}
    unix
       [Members OF PIPE, PID
          [newpipe:PIPE]
          [child[Init[exec(arg:string Array)]]:PID]
          [mk_stdin(x:PIPE)]
          [mk_n(x:PIPE)]
          [mk_stderr(x:PIPE)]
          [source OF W
             (x:PIPE)
             [Access(in:Input, __:Input):W]:W]
          [dest OF W
             (x:PIPE)
             [Access(out:Output, __:Output):W]:W]
          [close(x:PIPE)]
          [kill(p:PID)]
          [await(id:PID):int]
          [await_all]
          [run(Cmd:string Array)
             [Con OF IO
                [<(NONE,string) : IO]
                [>(NONE,string) : IO]
                [>>(NONE,string) : IO]
                [<(NONE,PIPE) : IO]
                [>(NONE,PIPE) : IO]
                [<(NONE,Input) : IO]
                [>(NONE,Output) : IO]:IO List]]]
\end{verbatim}
\end{Box}
The names \texttt{PIPE} and \texttt{PID} are member names of types, i.e.~fresh types that should not be confused with others, not even the same in another \texttt{unix}-application.

Member \texttt{run} is an example of a complex member that can be used class-like. It is used to invoke programs in separate processes and connect them in various ways, e.g.~with pipes or to files.  In terms of low-level primitives, the second argument of \texttt{run} has to describe how file descriptors are used as ports for interconnections.

A typical programming error with UNIX which prevents termination of a process control program, is failure to close a pipe for which a controlled process expects an end-of-file condition to arise. A designer may strive to define a complex operation so that disciplines to guard against such errors are enforced.

\subsection{\hhref[ref:contents]{Sorting}}

It has been claimed that operations may be class like, and that the notion of classes may be generalised and have more than a single argument where members are introduced. It seems not to be widely needed but please note:
\begin{itemize}
\item
Some languages have a predefined operation \texttt{sort} that distinguish an input and an output phase.  An operation can be defined with an argument for each phase, one with an \texttt{put}-operation, another with an operation to receive elements after ordering:
\begin{quote}
\begin{verbatim}
sort OF T,W  
   [ InOrder [x:T][y:T]:int ] 
   [ Send [put[X:T]] ] 
   [ Receive [all[Body[x:T]]]:W] :W
\end{verbatim}
\end{quote}
The first argument allows users to specify an ordering, e.g.~as
\verb|x<y|. 
\item
There is a nice symmetry between the definition and application parts of an internal definition.  So an internal definition can itself be considered an operation with two arguments with different `member names'.  
\item
Class-like operations are defined in terms of an argument name that defines meanings of members by an application of a name to many arguments.  An example is the \texttt{Scope}-argument of operation \texttt{var}.  A virtual internal definition would use the pattern
(with labels as reminders before each argument):
\begin{quote}
\begin{verbatim}
Scope __ { ... }   _ { ... }  asg { ... }
\end{verbatim}
\end{quote}
\end{itemize}

% \section{\hhref[ref:contents]{Declarations and syntactic coercion}}

\section{\hhref[ref:contents]{Conclusions}}

The verb `to prove' is transitive, but it is possible to conceive the noun `proof' independently of an object as required by the verb.  A proof can be understood as a pattern of references to individual rules of a formal logic.  In that sense a program can be seen as a proof provided each construct used in the composition of a program has a unique proof-rule associated with it.  A foundation to build a programming language with just such operations has been presented.

A benefit of programs as proofs is that not all programmers need to master the logic aspects of programs completely.  One may still acknowledge programmers work as proofs in a substantial sense.  Designers should, of course, be more concerned with the rules for application of their operations, i.e.~their adequateness as axioms of a logic.
\nopagebreak

Another benefit might be the implicit enforcement of a programming discipline that is indirectly influenced by formal logic, much the same as discussions in daily life may benefit from rules justified by formal logic.

Howard languages have operations that identify axioms, even if not explicitly expressed in a description.  Hence one might extend the Curry-Howard correspondence by considering a specific Howard language as a theory. 

A traditional proof of a program is merely a proof check that the given proof is effective in a given context.  Failure might stem from an incomplete case analysis, and that might lead to useful feed-back to programmers.

A programming language with constructs that all have signatures and hence possibly an internal definition rightly deserves to be called modular. New constructs can be implemented independently and added to extend existing languages.  So languages and their semantics can be built incrementally.

With the very broad notion of operations, it is easy to develop components by separate teams, if only they agree on component signatures.  The intended semantics can and should be negotiated between teams from the outset and revised during development with cost estimates of changes.  

Concepts about programming languages, semantics, and logic in the sense of transformation rules stem from previous work on a statement-oriented approach to data abstraction \cite{Steensgaard-Madsen:1981:SOA}.  For programming practice it seems to be a realistic realisation of programs as proofs.  

Experimental tools have been developed to support programming language design, interpreter construction, and formalised documentation of semantics of operations.  The tools also help to incrementally implement and install new operations with given signatures as predefined entities.

The programs-as-proofs aspect of the Curry-Howard correspondence is corroborated by this work.  The propositions-as-types aspect only to a lesser degree.  One can illustrate this with an expression outline

\[\TE[\theta:T]{P}R_{\theta}\]
making it clear that $T$ and $R_\theta$ have different r\^oles.  Although it might be tempting to have $T$ express the set of primes (or even subsets of primes) it seems simpler to keep the notions apart.  Not the least with regard to acceptance by programmers.

Mathematicans rarely rely on fully formal proofs, but formally state their goals.  So perhaps one might say: programmers rarely rely on formally stated goals, but formally state their proofs as programs.

% \vfill\newpage
% \addtocounter{section}{1}
% \section{\hhref[ref:contents]{References}}

\bibliographystyle{plain}
\begin{ebook}
\bibliography{jsm,types,temp,books}
\end{ebook}
\begin{paper}
\bibliography{plast}
\end{paper}

\end{document}